\documentclass[10pt,a4paper,aps,prl,twocolumn,showpacs,superscriptaddress,groupedaddress]{revtex4-1} 
\usepackage[english]{babel}
\usepackage{url} 
\usepackage{hyperref}
\usepackage{amsmath}
\usepackage{amsfonts}
\usepackage{slashed} 
\usepackage{graphicx} 
\usepackage{color}
\usepackage{bm}
\usepackage{textcomp}

\usepackage{pdfpages}
\makeatletter
\AtBeginDocument{\let\LS@rot\@undefined}
\makeatother

\newcommand{\bra}[1]{\langle #1 \vert}
\newcommand{\ket}[1]{\vert #1 \rangle}
\newcommand{\braket}[2]{\langle #1 \vert #2 \rangle}
\newcommand{\corr}[1]{\textcolor{black}{#1}}


\begin{document}
\allowdisplaybreaks

\title{Quantum Limitation to the Coherent Emission of Accelerated Charges}

\author{A. Angioi}
\affiliation{Max-Planck-Institut f\"ur Kernphysik, Saupfercheckweg 1, 69117 Heidelberg, Germany}
\author{A. Di Piazza} \email{dipiazza@mpi-hd.mpg.de}
\affiliation{Max-Planck-Institut f\"ur Kernphysik, Saupfercheckweg 1, 69117 Heidelberg, Germany}

\date{\today}
\begin{abstract}
Accelerated charges emit electromagnetic radiation. According to
classical electrodynamics if the charges move along sufficiently close 
trajectories they emit coherently, i.e., their emitted energy
scales quadratically with their number rather than
linearly. By investigating the emission by a two-electron wave 
packet in the presence of an electromagnetic plane wave within strong-field QED, 
we show that quantum effects deteriorate the coherence predicted
by classical electrodynamics even if the typical quantum nonlinearity 
parameter of the system is much smaller than unity. We explain this result by 
observing that coherence effects are also controlled by a new
quantum parameter which relates the recoil undergone by the
electron with the width of its wave packet in momentum space.
\end{abstract}

\maketitle

Optical laser pulses with intensities of the order of $10^{22}\, \text{W/cm}^2$ have been already
achieved~\cite{Yanovsky_2008} and intensities of the order of $10^{24}\, \text{W/cm}^2$ 
are envisaged~\cite{eli, xcels}. At such high intensities the interaction between
the laser field and an electron (mass $m$ and charge $e<0$) is highly-nonlinear
and electrodynamical processes involving electrons/positrons occur with the exchange of several
photons between the laser field and electrons/positrons themselves~\cite{ritus85,dipiazzareview}. This has also primed a surge of interest in testing QED in the so-called ``strong-field'' regime where the background field intensity is effectively of the order of $\mathcal{I}_{cr}=4.6\times 10^{29}\,\text{W/cm}^2$, 
corresponding to the electric field $\mathcal{E}_{cr}=m^2c^3/\hbar|e|=1.3\times 10^{16}\,\text{V/cm}$~\cite{dipiazzareview}. Due to the Lorentz invariance of the theory, in fact, strong-field QED can be effectively probed at laser intensities $\mathcal{I}\ll \mathcal{I}_{cr}$ by employing 
ultrarelativistic electron beams with correspondingly high energies 
$\varepsilon\sim mc^2\sqrt{\mathcal{I}_{cr}/\mathcal{I}}\gg mc^2$~\cite{dipiazzareview}.
Indeed, electron beams with energies beyond $1\, \text{GeV}$ have been already
produced both via conventional \cite{PDG_2017} and laser-based 
accelerators \cite{Leemans_2014}. One of the fundamental 
processes which can be exploited to test strong-field QED is Nonlinear 
Single Compton Scattering (NSCS), where an electron traveling inside a laser field 
exchanges multiple photons with the laser field itself while also emitting 
a single, non-laser photon. NSCS has been studied in the presence of a monochromatic
plane wave~\cite{Goldman_1964,kibble64, nikishov64, fried64, ritus85, ivanov04, harvey09,corson11b, wistisen14}, of a pulsed plane wave~\cite{boca09,heinzl10,boca11,felix11,seipt11,boca11,dinu12,dinu12b,dinu13,krajevska14,titov2014,angioi16}, and of a space-time-focused laser beam~\cite{dipiazza17}
(see also \cite{dipiazza14,dipiazza15,ilderton17,ilderton17b}). In \cite{kibble64, nikishov64, fried64, ritus85, ivanov04,harvey09,wistisen14,boca09,heinzl10,boca11,felix11,seipt11,boca11,dinu12,dinu12b,dinu13,krajevska14,titov2014} an incoming electron in a plane wave with a definite momentum was investigated, whereas in \cite{corson11b,angioi16} NSCS by a localized electron wave packet was studied. In all these works, the radiation emitted by a single electron has been considered, such that coherence effects in the nonlinear emission by several electrons have never been investigated within strong-field QED.

In this Letter we explore the novel features in the quantum radiation spectrum 
brought about by considering two-electron wave packets properly anti-symmetrized
as an initial state. For a single electron with definite asymptotic four-momentum 
$p^{\mu}$ the quantum spectra tend to the classical ones if $\chi = (kp) \mathcal{E} / m \omega \mathcal{E}_{cr}\ll 1$ \cite{ritus85,dipiazzareview}. Here, $\mathcal{E}$ and $k^\mu=(\omega,\bm{k})$ are the laser field's amplitude and its central four-wave-vector, respectively (units with $\hbar = c = 1$ and $\alpha=e^2\approx 1/137$ are employed throughout and the metric tensor reads $\eta^{\mu\nu}=\text{diag}(+1,-1,-1,-1)$). Now, according to classical physics, if $N$ charges move inside a field along sufficiently close trajectories, the radiated energy can scale as $N^2$ (rather than $N$) up to arbitrarily high 
frequencies~\cite{klepikov85}. Below, we consider the paradigmatic case where the two electrons 
are characterized by the same initial distribution of momenta and thus 
by the same average quantum parameter $\chi'$. We show that at very different
size scales of the \corr{electrons' wave packet} quantum effects 
limit or completely suppress the coherence of the emission even for 
$\chi'\ll 1$, i.e., when single-particle classical and quantum spectra approximately coincide. We note that in general for an initial multi-particle state the condition $\chi'\ll 1$ is not sufficient to recover the 
classical limit. However, our results explicitly indicate that the 
intuitive implication that when every particle emits classically then 
the whole system does too is invalid. We explain this unexpected result 
by observing that the condition $\chi'\ll 1$ ensures 
that the typical emitted photon energies are much smaller than the common 
average energy of electron wave packets. However, coherence effects are 
also controlled by a new quantum parameter which relates the recoil undergone by 
the electron not with the average energy but with the width of its wave packet 
in momentum space. These coherence effects, which become even larger at $\chi'\sim 1$, allow for high-precision tests of the strong-field sector of QED at the level of quantum amplitudes\corr{, which} employ 
few-electron pulses in a well-controlled quantum state.

The laser field is assumed to be linearly polarized along the $x$ direction and to propagate along the $z$ direction. Within the plane-wave approximation, it can be described by the classical four-potential $\mathcal{A}^\mu_L(\phi)=(0,\bm{\mathcal{A}}^\mu_L(\phi))=\mathcal{A}^{\mu}\psi_L(\phi)$, where $\mathcal{A}^{\mu}=(0,-\mathcal{E}/\omega,0,0)$,
$\psi_L(\phi)$ is a smooth function with compact support and $\phi=(nx)$, with $n^{\mu}=k^{\mu}/\omega=(1,0,0,1)$. For the sake of definiteness, we set $\phi=0$ as the initial light-cone ``time'' and thus assume that $\psi_L(\phi)=0$ for $\phi\le 0$. The initial two-electron state is characterized by two definite spin quantum numbers $s_j$ ($j\in\{1,2\}$) and has the form
\begin{equation}
  \label{eq:totalstat}
  \ket{\Psi} = \frac{1}{\sqrt{\mathcal{N}}}\prod_{j=1}^2\bigg[
  \int\frac{ d^3p_j}{(2\pi)^3\sqrt{2 \varepsilon_j}}
  \rho_j(\bm{p}_j)a_{s_j}^\dagger(\bm{p}_j)\bigg]\ket{0}.
\end{equation}
Here, $\mathcal{N}$ is a normalization factor such that $\braket{\Psi}{\Psi}=1$, the operator $a^{\dagger}_{s_j}(\bm{p}_j)$ creates an electron with momentum $\bm{p}_j$ (energy $\varepsilon_j = \sqrt{m^2 + \bm{p}_j^2}$) and spin quantum number $s_j$, $\rho_j(\bm{p}_j)$ is an arbitrary square-integrable complex-valued function describing each initial electron momentum distribution, and $\ket{0}$ is the free vacuum state. From the anti-commutation relations $\{a_{s}(\bm{p}), a^\dagger_{s'}(\bm{p}')\} = (2\pi)^3 \delta^{(3)}(\bm{p} - \bm{p}')\, \delta_{ss'}$~\cite{peskin}, the normalization factor $\mathcal{N}$ turns out to have the form $\mathcal{N}=\mathcal{N}_{12} - \delta_{s_1s_2} \mathcal{N}_{21}$, with
\begin{equation}
\label{N}  
\mathcal{N}_{ij} = \int\frac{ d^3p_1}{(2\pi)^32\varepsilon_1}
  \frac{d^3p_2}{(2\pi)^3 2\varepsilon_2}
  \rho_1^*(\bm{p}_1)\rho_i(\bm{p}_1)\rho_2^*(\bm{p}_2)\rho_j(\bm{p}_2).
\end{equation}

If $c_{l'}^\dagger(\bm{k}')$ is the operator which creates a photon with momentum $\bm{k}'$ (energy $\omega'= \sqrt{\bm{k}^{\prime 2}})$ and polarization $l'$, the final state in NSCS has the form
\begin{equation}
\ket{\Psi'} = \sqrt{8\omega'\varepsilon'_1\varepsilon'_2}
\, c_{l'}^\dagger(\bm{k}') a_{s'_2}^\dagger(\bm{p}'_2) a_{s'_1}^\dagger(\bm{p}'_1)\ket{0},
\end{equation}
with $\varepsilon'_j = \sqrt{m^2 + \bm{p}_j^{\prime\,2}}$. The leading-order $S$-matrix element $S$ of NSCS within the Furry picture \cite{furry, berestetskii82} reads
\begin{equation}
S = -ie\int d^4x\bra{\Psi'} \bar{\Psi}(x)\gamma^\mu\Psi(x)A_\mu(x)\ket{\Psi},
\end{equation}  
where the Dirac field $\Psi(x)$ is expanded with respect to Volkov states (see \cite{furry,berestetskii82,ritus85} and the Supplemental Material (SM) \footnote{See SM for an expression of the positive-energy Volkov states, for other well-known textbook results, and for an estimate of the repulsive effects between the two electrons outside the laser field.}), where $\bar{\Psi}(x)=\Psi^{\dagger}(x)\gamma^0$, where $\gamma^\mu$ are the Dirac matrices, and where $A^\mu(x)$ is the quantized part of the electromagnetic field. Here, we neglect the interaction between the electrons as their dynamics is predominantly determined by the intense plane wave. 

\begin{figure}
  \includegraphics[width=\linewidth]{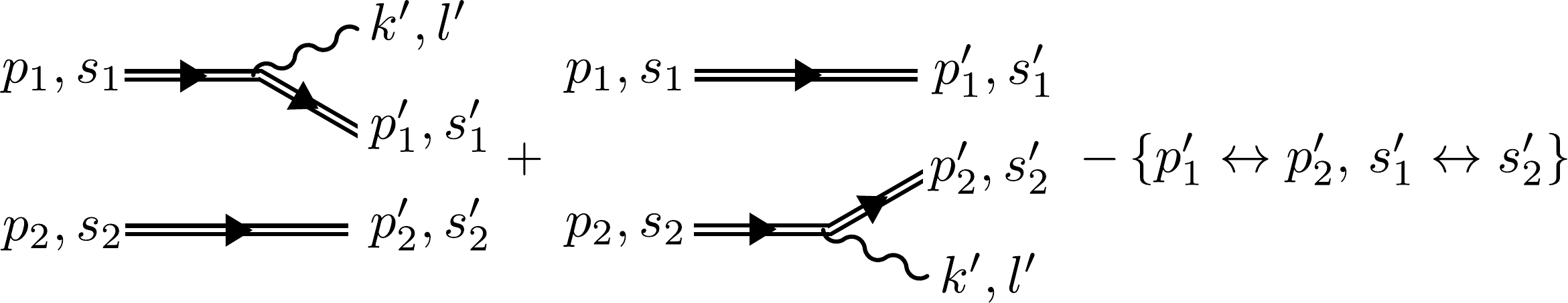}
  \caption{Leading-order Feynman diagrams of NSCS by two electrons. The double lines indicate Volkov states and the symbol $\{p'_1\leftrightarrow p'_2,s'_1\leftrightarrow s'_2\}$ indicates the exchange diagrams.
  }
  \label{fig:feyn}
\end{figure}

At the leading order of perturbation theory, only one of the two electrons emits a photon (see Fig.~\ref{fig:feyn}), the state of the other electron remaining unchanged.
Also, since the plane wave depends on the spacetime coordinates only via $\phi=t-z$, the amplitudes involving the photon emission include a three-dimensional Dirac delta-function, which enforces the conservation of the transverse ($\perp$) components ($x$- and $y$-components) and of the minus ($-$) component (time- minus $z$-component) of the four-momenta of the involved particles (see also the SM). Thus, by introducing the two on-shell four-momenta $q_j^\mu$ ($q_j^2  = m^2$) such that $\bm{q}_{j,\perp} = \bm{p}^{\prime}_{j,\perp} + \bm{k}'_{\perp}$ and $q_{j,-}=p'_{j,-}+k'_-$, i.e.,
\begin{equation}
\label{q_j}
q^{\mu}_j=p^{\prime\,\mu}_j+k^{\prime\,\mu}-\frac{(k'p'_j)}{p'_{j,-}+k'_-}n^{\mu},
\end{equation}
the amplitude $S$ can be written in the form $S=S_{12}-S_{21}$, where
\begin{equation}
\label{S_12}
S_{12} =\frac{e}{i}\sqrt{\frac{4 \pi}{\mathcal{N}}} \sum_{j=1}^2 \rho_j(\bm{q}_j)\rho_{j'}(\bm{p}'_{j'})\delta_{s'_{j'} s_{j'}}\frac{M_{s'_jl',s_j}(p'_j,k';q_j)}{2q_{j,-}},
\end{equation}
with $j'=3-j$, and where $S_{21}=S_{12}(p'_1\leftrightarrow p'_2,q_1\leftrightarrow q_2,s'_1\leftrightarrow s'_2)$. Here, we have introduced the reduced amplitude $M_{s'l',s}(p',k';p)$ characteristic of NSCS by a single electron with definite initial (final) four-momentum $p^{\mu}$ ($p^{\prime\,\mu}$) and spin quantum number $s$ ($s'$), which emits a photon with four-momentum $k'$ and polarization $l'$ (see, e.g., \cite{felix11} and the SM).

The emitted photon energy spectrum $dE_Q/ d \omega'$ of interest here reads
\begin{equation}
\label{spectrum}
  \frac{d E_Q}{ d \omega'} = \frac {\omega^{\prime\,2}}{8} \sum_{\{s_j s'_jl'\}} 
  \int \frac{ d \Omega'}{(2\pi)^32}\frac{ d^3 p'_1}
      {(2\pi)^3 2\varepsilon'_1}\frac{ d^3 p'_2}
      {(2\pi)^3 2\varepsilon'_2}\lvert S \rvert^2,
\end{equation}
where $\Omega'$ denotes the solid angle corresponding to $\bm{n}'=\bm{k'}/\omega'$. Note that if the electrons were distinguishable, the energy emission spectrum would have the same form as in Eq. (\ref{spectrum}), with the replacement $|S|^2\to 2(\mathcal{N}/\mathcal{N}_{12})|S_{12}|^2$.

In order to investigate the coherence properties of the emitted radiation, we consider the paradigmatic case in which the two electron wave packets in position space differ only by a translation by a vector $\bm{r}'$, i.e., $\rho_2(\bm{p}_2) = \rho_1(\bm{p}_2) \exp(-i \bm{p}_2\cdot\bm{r}')$, such that $|\rho_1(\bm{p})|^2=|\rho_2(\bm{p})|^2$. Also, without loss of generality we choose the function $\rho_1(\bm{p}_1)$ to be real and we denote it as $\rho(\bm{p}_1)$. 

Let us first study the classical energy spectrum $dE_C/ d \omega'$ emitted by two electrons in a plane wave with initial (at $t=0$) positions $\bm{r}'_1=\bm{0}$ and $\bm{r}'_2=\bm{r}'$, with $z'>0$, and four-momenta $p_j^{\prime\,\mu}=(\varepsilon'_j,\bm{p}'_j)$. Classical coherence effects in the emitted frequency $\omega'$ are controlled by the two phases $\omega'\Phi_j(\phi)$, with (see the SM)
\begin{equation}
\Phi_j(\phi)=\int_0^{\phi} d\phi'\frac{(n'p'_j(\phi'))}{p'_{j,-}}+\bigg[\frac{(n'p'_j)}{p'_{j,-}}\bm{n}-\bm{n}'\bigg]\cdot\bm{r}'_j.
\end{equation}
Here, $n^{\prime\,\mu}=k^{\prime\,\mu}/\omega'=(1,\bm{n}')$ or $\Phi_j(\phi)=\Phi_j(0)+n'_-\int_0^{\phi} d\phi'[m^2 +\bm{P}^{\prime\,2}_{j,\perp}(\phi')]/2p^{\prime\,2}_{j,-}$, where $\bm{P}'_{j,\perp}(\phi)=\bm{P}'_{j,\perp}-e\bm{\mathcal{A}}_{L,\perp}(\phi)$, with $\bm{P}'_{j,\perp}=\bm{p}'_{j,\perp}-p'_{j,-}\bm{n}'_{\perp}/n'_-$. Now, by indicating as $\varphi_T$ a measure of the total laser phase $\omega\phi_T$ where the electrons experience the strong field, an order-of-magnitude condition for the emitted radiation to be coherent is obtained by requiring that $\omega'\Delta\Phi(\phi_T)\lesssim \pi/5$ \footnote{This condition is obtained starting from the prototype function $g(\theta)=|1+\exp(i\theta)|^2$ and by stating that it shows a ``coherent'' behavior for $\theta<\theta^*$, where $\theta^*$ is such that $|g(\theta^*)-4|/4=0.1$, i.e. $\theta^*\approx \pi/5$}, with $\Delta\Phi(\phi_T)=|\Phi_2(\phi_T)-\Phi_1(\phi_T)|$ (the absolute value of the variation of an arbitrary quantity $f$ is indicated here and below as $\Delta f$). Now, we assume that the electrons have initial momenta (energies) of the same order of magnitude $\bm{p}'$ ($\varepsilon'$), and that are ultrarelativistic and initially counterpropagating with respect to the laser field ($p'_-/2\approx \varepsilon'\gg m$). By summing the moduli of all contributions to $\Delta\Phi(\phi_T)$, the above condition provides an upper limit $\omega'_C$ on the frequencies which are emitted coherently given by
\begin{equation}
  \omega'_C=
  \frac{2\pi\omega}{5n'_-\varphi_T}\left[
    \frac{\Delta\overline{\bm{P}^{\prime\,2}_{\perp}}}{4\varepsilon^{\prime\,2}}
    +
    \frac{\Delta\varepsilon'}{\varepsilon'}
    \frac{m^2+\overline{\bm{P}^{\prime\,2}_{\perp}}}{2\varepsilon^{\prime\,2}}+\frac{2\omega\Delta\Phi(0)}{n'_-\varphi_T}\right]^{-1},
  \label{eq:estimate}
\end{equation}
where $\overline{\bm{P}^{\prime\,2}_{\perp}}$ is the average value of $\bm{P}^{\prime\,2}_{\perp}(\phi)$ over $\phi_T$. It is physically clear that the larger the interaction time is and the larger the differences in the electrons' initial positions/momenta/energies are, the lower will be the highest frequency that can be emitted coherently. Notice that the quantity $\Delta\Phi(0)$ in Eq. (\ref{eq:estimate}) depends on the initial distance
$|\bm{r}'_2-\bm{r}'_1|$ of the two electrons.

Having in mind the quantum case where \corr{the electrons' momentum distributions} are given by $\rho^2(\bm{p}'_1)$ and $\rho^2(\bm{p}'_2)$, we consider now a classical ensemble of pairs of electrons, each pair being characterized by the electrons' initial positions $\bm{r}'_1=\bm{0}$ and $\bm{r}'_2=\bm{r}'$ and initial (and final) momenta $\bm{p}'_j$ distributed as $\rho^2(\bm{p}'_1)$ and $\rho^2(\bm{p}'_2)$. The corresponding average classical energy spectrum $\langle dE_C/ d \omega'\rangle$ reads
\begin{equation}
\left\langle\frac{d E_C}{d \omega'}\right\rangle=
  \int\frac{ d^3p'_1}{(2\pi)^32 \varepsilon'_1}
  \frac{ d^3p'_2}{(2\pi)^32 \varepsilon'_2}
  \frac{\rho^2(\bm{p}'_1)\rho^2(\bm{p}'_2)}{\mathcal{N}_{12}}\frac{dE_C}{ d \omega'}.
\label{eq:semicl}
\end{equation}
This expression can also be obtained from the quantum spectrum $dE_Q/ d \omega'$ in Eq.~(\ref{spectrum}) by neglecting the photon recoil in $\rho(\bm{q}_j)$, i.e., by approximating $\rho(\bm{q}_j)\approx\rho(\bm{p}'_j)$, but by keeping linear corrections due to the recoil in the phase of $\rho_2(\bm{q}_2)$. This, in fact, allows to reproduce the term $\Phi_2(0)$ from the difference $\bm{q}_2-\bm{p}'_2$ according to Eq.~(\ref{q_j}) after neglecting higher-than-linear recoil terms in it, which in turn describes the role of the wave packets' separation $\bm{r}'$. On the one hand, this implies that when the photon recoil is negligible, the classical constraint in Eq.~(\ref{eq:estimate}) also applies quantum mechanically. On the other hand, however, we will show below that the differences in the coherence properties of classical and quantum radiation precisely arise from the fact that the classical theory ignores the recoil in $\rho(\bm{q}_j)$. In fact, turning now to the quantum case, it is intuitively clear, as we have also ascertained in the numerical examples below, that the electrons' indistinguishability does not play a significant role here (the exchange term slightly reduces the emitted energy). Indeed, the exchange terms become important only when the two electrons have very similar final momenta (and the same final spin), which corresponds to a negligibly small region of the available final phase space. Thus, in order to study coherence effects, we focus on the interference term in $|S_{12}|^2$, which is proportional to the product $\rho(\bm{q}_1)\rho(\bm{p}'_1)\rho(\bm{p}'_2)\rho(\bm{q}_2)$ [see Eq. (\ref{S_12})]. In analogy with the classical case, we indicate as $\bm{p}'$ the average momentum of both electron distributions, corresponding to the on-shell four-momentum $p^{\prime\,\mu}=(\varepsilon',\bm{p}')=(\sqrt{m^2+\bm{p}^{\prime\,2}},\bm{p}')$, and as $\bm{\sigma}_{\bm{p}'}$ the three-dimensional width. As it is clear from Eq. (\ref{q_j}), the difference between the momenta $\bm{p}'_j$ and $\bm{q}_j$ is due to the photon recoil. Thus, if the latter is so large that $|p'_{j,i}-q_{j,i}|\gg \sigma_{p'_i}$ for any $i\in\{x,y,z\}$, the interference term will be suppressed because the functions $\rho(\bm{q}_j)=\rho(\bm{q}_j(\bm{p}'_j))$ [see Eq. (\ref{q_j})] and $\rho(\bm{p}'_j)$ cannot be both significantly different from zero for the same $\bm{p}'_j$. As a result, the interference term in $|S_{12}|^2$ will be suppressed and the radiation with frequency $\omega'\gtrsim\omega'_Q=\min_i\{\sigma_{p'_i}/|\cos\theta'_i|\}$, with $\theta'_i$ being the angle between $\bm{k}'$ and the $i$th axis, will be incoherent [the last term in Eq. (\ref{q_j}) has been neglected, which is a good approximation in most situations of interest]. An invariant parameter $\tilde{\chi}(k')$ characterizing the quantum coherence of the emitted radiation with four-momentum $k^{\prime\,\mu}$ can be defined by introducing the average $\langle\langle\cdot\rangle\rangle$ with respect to the distribution $\rho^2(\bm{p}')/\mathcal{N}_0$, with $\mathcal{N}_0=\int d^3p'(2\pi)^{-3}\rho^2(\bm{p}')/2\varepsilon'$: $\tilde{\chi}(k')=\sqrt{k^{\prime\,\mu}T^{-1}_{\mu\nu}k^{\prime\,\nu}}$, where $(T^{-1})^{\mu\nu}$ is the inverse of the  positive-definite, symmetric covariance tensor $T^{\mu\nu}=\langle\langle p^{\prime\,\mu}p^{\prime\,\nu}\rangle\rangle-\langle\langle p^{\prime\,\mu}\rangle\rangle\langle\langle p^{\prime\,\nu}\rangle\rangle$. The matrix $T^{\mu\nu}$ can be diagonalized by means of a Lorentz transformation $\Lambda^{\star}$ \cite{Blaettel_1989}. If the resulting diagonal matrix $T^{\star}=\Lambda^{\star}T\Lambda^{\star\,t}$ reads $T^{\star\,\mu\nu}=\text{diag}(\Sigma^2_{\varepsilon'},\Sigma^2_{p'_x},\Sigma^2_{p'_y},\Sigma^2_{p'_z})$, then $\Sigma^2_{\varepsilon'}$ and  $\Sigma^2_{p'_i}$ are the variances of the energy and of the $i$-th component of the momentum distribution in that frame. Thus, \corr{we have}
\begin{equation}
\tilde{\chi}(k')=\sqrt{\frac{\omega^{\prime\,\star\,2}}{\Sigma^2_{\varepsilon'}}+\frac{k_x^{\prime\,\star\,2}}{\Sigma^2_{p'_x}}+\frac{k_y^{\prime\,\star\,2}}{\Sigma^2_{p'_y}}+\frac{k_z^{\prime\,\star\,2}}{\Sigma^2_{p'_z}}},
\end{equation}
where $k^{\star\,\mu}=(\omega^{\star},\bm{k}^{\star})$ is the emitted photon four-momentum
in that frame. If $\tilde{\chi}(k')<1$ the coherence of the radiation with $k^{\prime\,\mu}$ is not deteriorated by quantum effects. 
  
The additional quantum restriction to the coherent emission of radiation is qualitatively different from the classical one and it can be related to the particles' ``kinematic'' indistinguishability. In fact, depending on the width of the electron wave packets, even a perfect knowledge of the final momenta of the two electrons and of the emitted photon combined with the momentum conservation laws does not allow to know with certainty which electron has emitted. In this respect, different momentum components of the two-electron wave packet $\ket{\Psi}$ constructively interfere enhancing the radiation probability. This is in striking contrast with the case of an incoming single electron, where, indeed, the conservation laws allow to determine the initial momentum of the electron once the final electron and photon momenta are known, implying that the emission spectrum is given by the incoherent sum of the emissions spectra corresponding to each momentum component of the wave packet \cite{corson11b,angioi16}.

\begin{figure}
    \includegraphics[width=\linewidth]{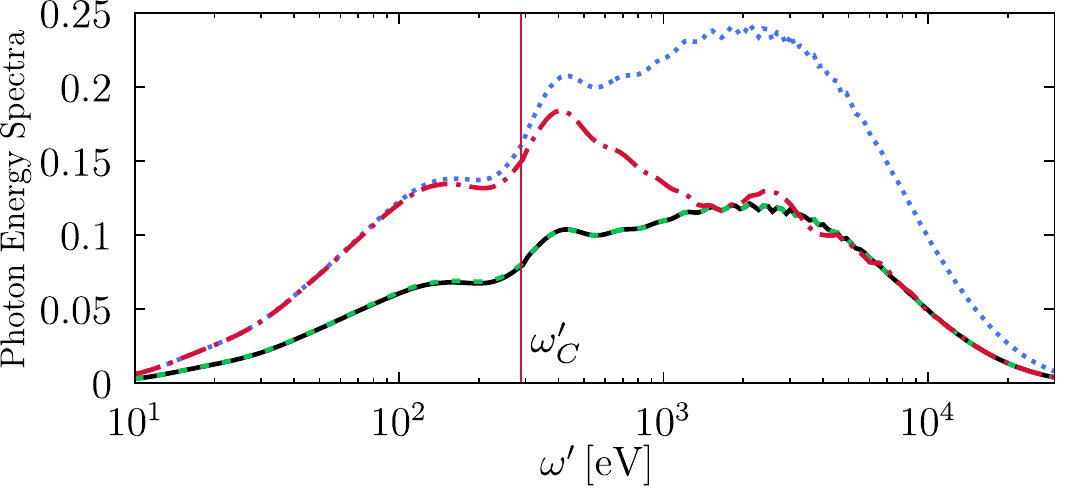}
    \caption{Energy spectra $dE_Q/ d \omega'$ (solid black line)
    and $\langle dE_C/d\omega'\rangle$ (dash-dotted red line) for numerical
      parameters given in the text and the single-electron spectrum multiplied 
      by two (dashed green line) and by four (dotted blue line).}
    \label{fig2}
  \end{figure}

Below, we show that the quantum restriction to the coherence of the emission can be essentially more restrictive than the classical one even if the average quantum parameter $\chi'=(kp')\mathcal{E}/m\omega\mathcal{E}_{cr}$ of the two wave packets is much smaller than unity. In Fig.~\ref{fig2} we compare the full quantum spectrum $dE_Q/d\omega'$ from Eq. (\ref{spectrum}) (solid black line) with the classical spectrum $\langle dE_C/ d \omega'\rangle$ from Eq. (\ref{eq:semicl}) (dash-dotted red line).
As reference, we also show the single-electron spectra multiplied by two (dashed green line) and by four (dotted blue line), which are the same for the two electrons~\cite{angioi16}. Concerning the electrons, we have set $\bm{r}' = (10^{-2},10^{-2} ,10^{-3}) \,  \text{eV}^{-1} \approx (2,2,0.2) \, \text{nm}$ and $\rho^2(\bm{p}'_j)/2\varepsilon'_j$ to be a normalized Gaussian function, with average momentum $\bm{p}' = (0,0,-10\, \text{MeV})$, transverse standard deviation $\sigma_{p'_{\perp}} =\sigma_{p'_x} = \sigma_{p'_y}= 31 \,\text{eV}$ and longitudinal standard deviation $\sigma_{p'_{\parallel}}=\sigma_{p'_z}= 0.62 \, \text{eV}$. Concerning the plane wave, we have set $\omega = 1.55 \, \text{eV}$, $\mathcal{I} = 10^{20}\, \text{W}/\text{cm}^2$, and $\psi_L(\phi) = \sin^4(\omega \phi/4)\sin(\omega \phi)$ for $0\le \omega \phi \le 4\pi$ and zero elsewhere, such that $\chi' \approx 0.002$. Fig.~\ref{fig2} shows that the classical spectrum is coherent up to a given frequency, that can be calculated with  Eq. (\ref{eq:estimate}); for this estimate we choose $\bm{n}'\sim -(m\xi/2\varepsilon',0,1)$, with $\xi=|e|\mathcal{E}/m\omega = 5$, as a typical observation direction where the average radiated energy is large \cite{felix11}. We estimate the variations $\Delta \bm{p}'_{\perp}$ and $\Delta\varepsilon'\approx\Delta p'_{\parallel}$ entering Eq.~(\ref{eq:estimate}) via the standard deviations $\sigma_{p'_{\perp}}$ and $\sigma_{p'_{\parallel}}$, respectively. By also estimating $\varphi_T\sim 2\pi$ as the effective phase where the laser field is  strong, we find from Eq.~(\ref{eq:estimate}) that $\omega'_C \approx 278 \,\text{eV}$, in good agreement with Fig.~\ref{fig2} (see the red vertical line). The quantum spectrum (solid black line) is incoherent over the whole range shown in Fig.~\ref{fig2} because, by estimating $|\bm{k}'_{\perp}|\sim \omega'm\xi/\varepsilon'$, we obtain that $\omega'_Q\sim\min\{\sigma_{p'_{\perp}}\varepsilon'/m\xi,\sigma_{p'_{\parallel}}\}$, which corresponds to $\omega'_Q= \sigma_{p'_{\parallel}} =  0.62 \,\text{eV}$. Thus, even if classical arguments would predict coherent emission until $\omega'_C$, the lower bound $\omega'_Q$ given by quantum mechanics, being orders of magnitude smaller, dominates. According to the estimations in the SM, the Coulomb repulsion between the two electrons before they enter the laser field can be neglected for the above numerical parameters.

  \begin{figure}
    \includegraphics[width=\linewidth]{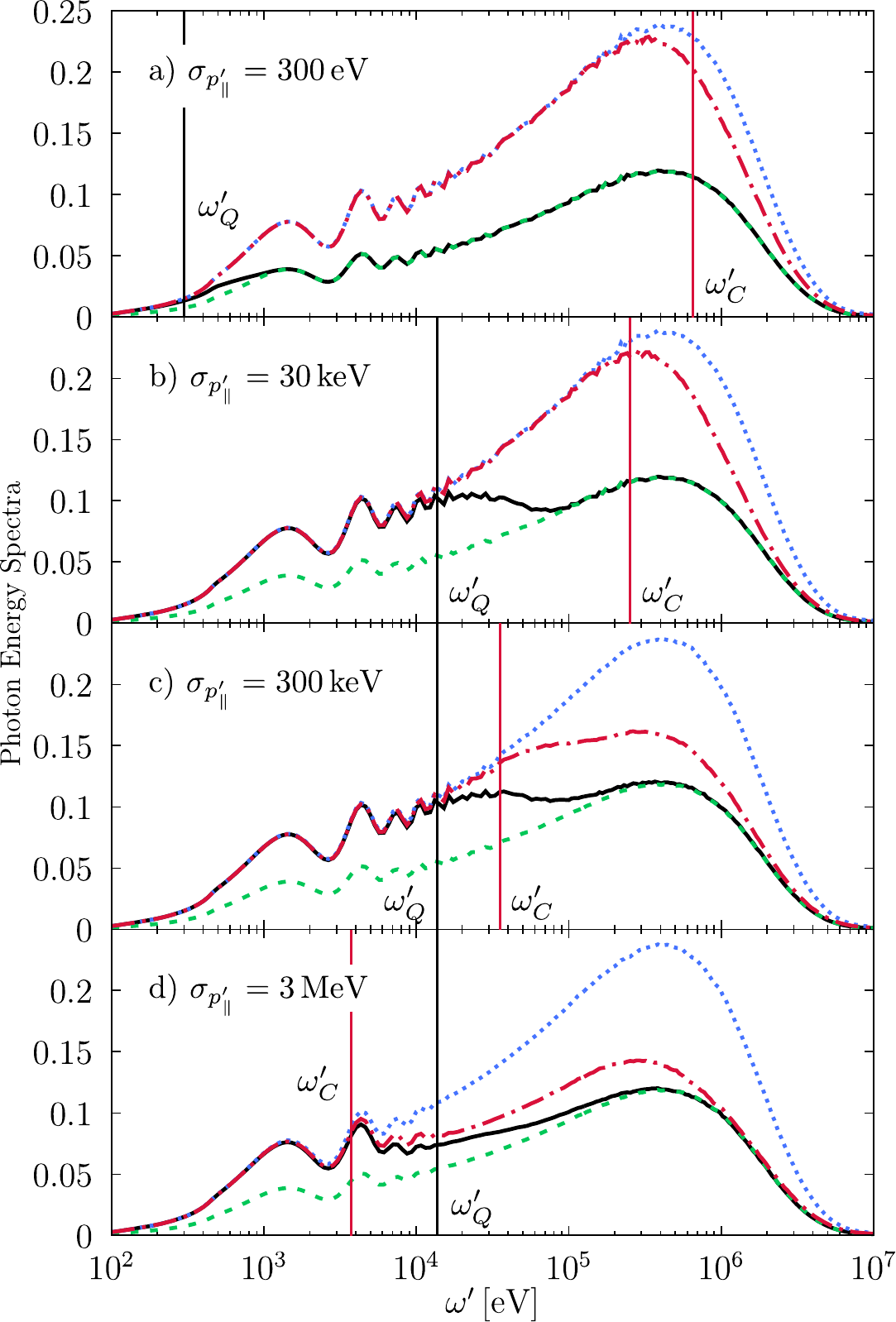}
    \caption{Energy spectra for numerical parameters given in the text. 
    The meaning of each line is the same as Fig.~\ref{fig2}.}
    \label{fig3} 
  \end{figure}
In Fig.~\ref{fig3} we provide a compact visualization of the interplay between the classical and quantum limits on coherent emission. In order to show all the effects we mentioned in a single graph without changing multiple numerical parameters, we have fixed them at the same values of Fig.~\ref{fig2}, except that $\bm{p}' = (0,0,-100\, \text{MeV})$, $\sigma_{p'_{\perp}} = 1 \,\text{keV}$, $\mathcal{I} = 1.2 \times 10^{21}\,\text{W}/\text{cm}^2$ ($\chi' \approx 0.02$), $\bm{r}' = (0,10^{-4},10^{-7})\,\text{eV}^{-1}$=(0,20,0.02)\,\text{pm}, and $\sigma_{p'_{\parallel}}$ is varying in each panel. The largest difference between classical and quantum results is observed in panel a), where the Coulomb repulsion is not expected to play a significant role, whereas the latter may significantly alter the average distance between the electrons before they enter the laser field in the case of panels b)-d) (see the SM).

The values of $\omega'_Q$, calculated in the same way as Fig.~\ref{fig2}, and of $\omega'_C$, estimated via Eq.~(\ref{eq:estimate}), are in reasonable agreement with the numerical results. In particular, it is interesting to observe that the quantum limit dominates in panels a)-c), where $\omega'_C>\omega'_Q$, and the classical limit takes over in panel d) where $\omega'_C<\omega'_Q$, where it applies to both the classical and the quantum spectrum. We notice that in any case interference effects always amount to an increase of the radiation yield, an enhancement effect essentially due to the reduced distance between the two wave packets.

The properties of single-electron pulses with energies of the order of $0.1\, \text{MeV}$
and attosecond duration are already being exploited experimentally in order to perform high-precision microscopy (see~\cite{baum13,kealhofer16,morimoto18,morimoto18b}), and control schemes for electrons of MeV energy have been
demonstrated recently~\cite{curry18}. Moreover, recent theoretical studies indicate the feasibility of generating arbitrarily-delayed single-electron wave packets with GeV energies \cite{Krajewska_2017}.
The extension of these techniques to few-electron wave packets seems possible~\cite{BaumPC}, for instance by combining two single-electron pulses with the methods of~\cite{kealhofer16,morimoto18, morimoto18b} or via an ultracold gas source~\cite{claessens05,geer09,franssen17}, where the electrons are already highly correlated
from the beginning.
Our results suggest that the development of similar techniques at higher energies would have important applications also in fundamental strong-field physics. By reversing the argument, we can also say that the NSCS spectra as calculated here can be exploited, provided a detailed knowledge of the laser pulse, as a diagnostic tool for two- or few-electron high-energy pulses.

%
\begin{acknowledgments}
The authors would like to acknowledge P. Baum, J. Evers, and K. Z. Hatsagortsyan for helpful discussions.
A. A. is also thankful to S. Bragin, S. Castrignano, S. M. Cavaletto, and O. D. Skoromnik
for the feedback they provided during the development of the ideas contained in this paper.
\end{acknowledgments}


%

\clearpage



\section*{Supplementary material}

# Quantum Limitation to the Coherent Emission of Accelerated Charges: Supplemental Material

A. Angioi and A. Di Piazza*

*Max-Planck-Institut für Kernphysik, Saupfercheckweg 1, 69117 Heidelberg, Germany*

In the present Supplemental Material we provide some results that we did not include in the main text because they are widely known in the literature (see e.g. [1–3]), and an estimate of the average strength of the Coulomb repulsion in the two-electron state considered in the main text. For the sake of clarity the numbers of the equations here contain "SM" in addition, such that the usual numbering (without SM) refers to the equations in the main text.

## VOLKOV STATES

As we mentioned in the main text, our work is carried out in the Furry picture [2]. Thus, the fermion field $\Psi(x)$ in Eq. (4) needs to be expanded by employing "dressed" electron states, i.e., the solutions of the Dirac equation in the background field. In our case the background field is a linearly-polarized plane wave described by the four-vector potential $\mathcal{A}_L^\mu(\phi) = \mathcal{A}^\mu \psi_L(\phi)$, where $\psi_L(\phi)$ is a smooth function with compact support, with $\phi = (nx) = t - z$, i.e., $n^\mu = (1, 0, 0, 1)$ [see also the main text above Eq. (1) for additional details], and the resulting solutions of the Dirac equation are the so-called Volkov states [2]. By setting $\phi = 0$ as the initial light-cone "time" and by correspondingly assuming that $\psi_L(\phi) \equiv 0$ for $\phi \leq 0$, the positive-energy Volkov state of an electron with four-momentum $p^\mu$ and a spin quantum number $s$ at $\phi = 0$ reads

$$\psi_{ps}(x) = \left[ 1 + \frac{e}{2p_-} \not{n} \not{\mathcal{A}}_L(\phi) \right] u_{ps} e^{iS_p(x)}. \quad \text{(SM.1)}$$

Here, we have introduced the minus-component $p_- = (np)$, the "slash" notation $\not{v} = \gamma^\mu v_\mu$ for an arbitrary four-vector $v^\mu$, with $\gamma^\mu$ being the Dirac gamma-matrices, the constant bispinor $u_{ps}$ fulfilling $(\not{p} - m) u_{ps} = 0$ and normalized as $\bar{u}_{ps} u_{ps} = 2m$ [2], and the classical action $S_p(x)$ of the electron in the plane wave

$$S_p(x) = -(px) - \int_0^\phi d\phi' \left[ \frac{e(p\mathcal{A}_L(\phi'))}{p_-} - \frac{e^2 \mathcal{A}_L^2(\phi')}{2p_-} \right]. \quad \text{(SM.2)}$$

It is possible to associate the average four-current

$$j_p^\mu(\phi) = \frac{\bar{\psi}_{ps}(x) \gamma^\mu \psi_{ps}(x)}{\bar{\psi}_{ps}(x) \psi_{ps}(x)} \quad \text{(SM.3)}$$

to the Volkov state $\psi_{ps}(x)$ and it turns out that $j_p^\mu(\phi) = p^\mu(\phi)/m$, where $p^\mu(\phi) = (\varepsilon(\phi), \boldsymbol{p}(\phi))$ is the solution of the Lorentz equation in the plane wave with initial four-momentum $p^\mu$:

$$p^\mu(\phi) = p^\mu - e\mathcal{A}_L^\mu(\phi) + \left[ \frac{e(p\mathcal{A}_L(\phi))}{p_-} - \frac{e^2 \mathcal{A}_L^2(\phi)}{2p_-} \right] n^\mu. \quad \text{(SM.4)}$$

## NONLINEAR SINGLE COMPTON SCATTERING

The leading order $S$-matrix element $S_{fi}$ of nonlinear single Compton scattering by an electron with initial (final) four-momentum $p^\mu$ ($p'^\mu$) and spin quantum number $s$ ($s'$) is given within the Furry picture by

$$S_{fi} = -ie\sqrt{4\pi} \int d^4x \, \bar{\psi}_{p's'}(x) \not{\epsilon}_{l'}'^* e^{i(k'x)} \psi_{ps}(x), \quad \text{(SM.5)}$$

where $k'^\mu$ is the four-momentum of the emitted photon and $\epsilon_{l'}'^\mu$ its polarization four-vector, with $l' \in \{1, 2\}$. Since the integrand in Eq. (SM.5) depends non-trivially only on $\phi = t - z$, the remaining three spacetime integrals provide corresponding momenta delta-functions and $S_{fi}$ can be written in the form

$$S_{fi} = -ie\sqrt{4\pi}(2\pi)^3 \delta^{(x,y,-)}(p - k' - p') M_{s'l',s}(p', k'; p). \quad \text{(SM.6)}$$

By recalling the matrix structure of Volkov states [see Eq. (SM.1)], the reduced amplitude $M_{s'l',s}(p', k'; p)$ can be written as $M_{s'l',s}(p', k'; p) = \bar{u}_{p's'} \tilde{M}_{l'}(p', k'; p) u_{ps}$, with [3]

$$\tilde{M}_{l'}(p', k'; p) = \not{\epsilon}_{l'}'^* f_0 + e \left( \frac{\mathcal{A} \not{n} \not{\epsilon}_{l'}'^*}{2p'_-} + \frac{\not{\epsilon}_{l'}'^* \not{n} \mathcal{A}}{2p_-} \right) f_1 \\ - \frac{e^2 \mathcal{A}^2 \epsilon_{l',-}'^*}{2p_- p'_-} \not{n} f_2, \quad \text{(SM.7)}$$

where

$$f_b = \int d\phi \, \psi_L^b(\phi) e^{i \int_0^\phi d\phi' [a_0 + a_1 \psi_L(\phi') + a_2 \psi_L^2(\phi')]}, \quad \text{(SM.8)}$$

with $b \in \{0, 1, 2\}$ and

$$a_0 = \frac{(k'p)}{p'_-}, \tag{SM.9}$$

$$a_1 = \frac{e(p'\mathcal{A})}{p'_-} - \frac{e(p\mathcal{A})}{p_-}, \tag{SM.10}$$

$$a_2 = -\frac{e^2\mathcal{A}^2}{2}\frac{k'_-}{p_-p'_-}. \tag{SM.11}$$

Note that, by either enforcing gauge invariance or by integrating it by parts, the formally divergent integral $f_0$ can be regularized according to the identity $f_0 = -(a_1 f_1 + a_2 f_2)/a_0$ [3].

## CLASSICAL LIMIT

In order to make the classical limit mentioned below Eq. (10) more transparent, we notice that in the single-electron case considered here:

1. by neglecting the recoil in Eq. (SM.7), i.e., by approximating $p'_- \approx p_-$ there and $u_{p's'} \approx u_{ps'}$, the pre-exponential of the matrix $M_{s'l',s}(p', k'; p)$ reduces to $2(p(\phi)\epsilon'_{l'})\delta_{ss'}$;

2. by neglecting terms higher than linear in the photon recoil in the coefficients $a_0$, $a_1$ and $a_2$, i.e., by approximating $a_0 \approx (k'p)/p_-$, $a_1 \approx e[p_-(k'\mathcal{A}) - k'_-(p\mathcal{A})]/p_-^2$, and $a_2 \approx -e^2\mathcal{A}^2 k'_-/2p_-^2$, the phase of the integrands in the functions $f_b$ coincides with the corresponding classical phase in Eq. (8) (for an electron in the origin at $t = 0$).

## DERIVATION OF THE CLASSICAL CUT-OFF

Let us start from the classical energy spectrum $dE_C/d\omega'$ emitted by two electrons in a plane wave [1, 4]

$$\frac{dE_C}{d\omega'} = \frac{e^2\omega'^2}{4\pi^2}\int d\Omega' \left| \sum_{j=1}^{2}\int d\phi \frac{p_j'^{\mu}(\phi)}{p'_{j,-}}e^{i\omega'\Phi_j(\phi)} \right|^2. \tag{SM.12}$$

Here, for the sake of notational convenience in relation with the quantum case, we have indicated as $p_j'^{\mu}(\phi) = (\varepsilon_j'(\phi), \boldsymbol{p}_j'(\phi))$ the electrons' four-momenta in the plane wave [2]

$$p_j'^{\mu}(\phi) = p_j'^{\mu} - e\mathcal{A}_L^{\mu}(\phi) + \left[ \frac{e(p_j'\mathcal{A}_L(\phi))}{p'_{j,-}} - \frac{e^2\mathcal{A}_L^2(\phi)}{2p'_{j,-}} \right]n^{\mu}, \tag{SM.13}$$

with initial (at $t = 0$) four-momenta $p_j'^{\mu} = (\varepsilon_j', \boldsymbol{p}_j')$. Also, by labeling as "1" the electron which first enters the plane wave and by setting the origin of the coordinate system at the corresponding entering point, the initial positions

of the electrons are $\boldsymbol{r}_1' = \boldsymbol{0}$ and $\boldsymbol{r}_2' = \boldsymbol{r}'$, with $z' > 0$. The quantity $\Phi_j(\phi)$ in Eq. (SM.12) thus reads

$$\Phi_j(\phi) = \int_0^{\phi}d\phi' \frac{(n'p_j'(\phi'))}{p'_{j,-}} + \left[ \frac{(n'p_j')}{p'_{j,-}}\boldsymbol{n} - \boldsymbol{n}' \right] \cdot \boldsymbol{r}_j', \tag{SM.14}$$

with $n'^{\mu} = k'^{\mu}/\omega' = (1, \boldsymbol{n}')$ or $\Phi_j(\phi) = \Phi_j(0) + n'_- \int_0^{\phi}d\phi'[m^2 + \boldsymbol{P}_{j,\perp}'^2(\phi')]/2p'^2_{j,-}$, where $\boldsymbol{P}_{j,\perp}'(\phi) = \boldsymbol{P}_{j,\perp}' - e\mathcal{A}_{L,\perp}(\phi)$, with $\boldsymbol{P}_{j,\perp}' = \boldsymbol{p}_{j,\perp}' - p'_{j,-}\boldsymbol{n}'_{\perp}/n'_-$. Now, by indicating as $\varphi_T$ a measure of the total laser phase $\omega\varphi_T$ where the electrons experience the strong field, an order-of-magnitude condition for the emitted radiation to be coherent is obtained by requiring that $\omega'\Delta\Phi(\varphi_T) \lesssim \pi/5$. This condition is obtained starting from the prototype function $g(\theta) = |1 + \exp(i\theta)|^2$ and by stating that it shows a "coherent" behavior for $\theta < \theta^*$, where $\theta^*$ is such that $|g(\theta^*) - 4|/4 = 0.1$, i.e. $\theta^* \approx \pi/5$, with $\Delta\Phi(\varphi_T) = |\Phi_2(\varphi_T) - \Phi_1(\varphi_T)|$ (the absolute value of the variation of an arbitrary quantity $f$ is indicated here and below as $\Delta f$). Now, we assume that the electrons have initial momenta (energies) of the same order of magnitude $\boldsymbol{p}'$ $(\varepsilon')$, and that are ultrarelativistic and initially counterpropagating with respect to the laser field $(p'_-/2 \approx \varepsilon' \gg m)$. By summing the moduli of all contributions to $\Delta\Phi(\phi_T)$, the above condition provides an upper limit $\omega_C'$ on the frequencies which are emitted coherently given by

$$\omega_C' = \frac{2\pi\omega}{5n'_-\varphi_T}\left[ \frac{\Delta\overline{\boldsymbol{P}_{\perp}'^2}}{4\varepsilon'^2} + \frac{\Delta\varepsilon'}{\varepsilon'}\frac{m^2 + \overline{\boldsymbol{P}_{\perp}'^2}}{2\varepsilon'^2} + \frac{2\omega\Delta\Phi(0)}{n'_-\varphi_T} \right]^{-1}, \tag{SM.15}$$

where $\overline{\boldsymbol{P}_{\perp}'^2}$ is the average value of $\boldsymbol{P}_{\perp}'^2(\phi)$ over $\phi_T$.

## COULOMB REPULSION

In the calculations presented in the main text, we neglected the effects of the Coulomb interaction between the two electrons. On the one hand, this is a very good approximation when the electrons are inside the laser field, due to the overwhelming force due to the latter. On the other hand, however, the close proximity of the two charges could in principle render the Coulomb interaction non-negligible when the charges have not entered the laser field yet.

In order to estimate the strength of Coulomb interaction, since in the considered setup the electrons move along the same direction with the same average energy, one can start by evaluating the average inverse-squared distance between the two electrons in the frame of reference where the electrons are on average at rest. In this frame of reference, the longitudinal indeterminacy of the electrons in position space $\tilde{\sigma}_{\parallel}$ is dilated by the average Lorentz gamma factor $\gamma$ with respect to the longitudinal

indeterminacy in the lab frame $\sigma_\parallel = 1/2\sigma_{p'_\parallel}$, thus:

$$\tilde{\sigma}_\parallel = \frac{\gamma}{2\sigma_{p'_\parallel}}. \qquad (SM.16)$$

Moreover, also the longitudinal displacement between the two wave packets in the average rest frame $r'_\parallel$ is dilated: $\tilde{r}'_\parallel = \gamma r'_\parallel$. Instead, the transverse displacement and indeterminacy are not affected by the Lorentz boost, thus $\tilde{\sigma}_\perp = 1/2\sigma_{p'_\perp}$ and $\tilde{\boldsymbol{r}}'_\perp = \boldsymbol{r}'_\perp$.

We assume the electrons' wave packets to be Gaussian as in the main text such that the average value of the Coulomb force between the two electrons is given by

$$\langle \tilde{\boldsymbol{F}}_C \rangle = \alpha \int \frac{d^3r_1 d^3r_2}{(2\pi)^3 \tilde{\sigma}_\parallel^2 \tilde{\sigma}_\perp^4} \frac{\boldsymbol{r}_1 - \boldsymbol{r}_2}{|\boldsymbol{r}_1 - \boldsymbol{r}_2|^3} e^{-\sum_{l=1}^3 \frac{r_{1,l}^2 + (r_{2,l} - r'_l)^2}{2\tilde{\sigma}_l^2}}, \qquad (SM.17)$$

where $\alpha = e^2 \approx 1/137$ is the fine-structure constant. The integrals in Eq. (SM.17) can be evaluated numerically and for the parameters corresponding to Fig. 2, we obtain that $|\langle \tilde{\boldsymbol{F}}_C \rangle| \approx 3 \times 10^{-3}$ eV$^2$. The repulsive force is thus completely negligible with respect to the peak electric force that the laser field exerts on an electron in the same frame ($|e|\tilde{\mathcal{E}} \sim m\omega\gamma\xi = 1.6 \times 10^8$ eV$^2$). The same conclusion can be drawn for the numerical example corresponding to Fig. 3. Although this implies that the interaction between the electrons is negligible with respect to their interaction with the laser field, the interaction between the electrons before they enter the laser field could push them so far apart that there would be no interference in their radiation, neither classically nor quantum mechanically.

In order to determine the length scale at which this can occur, we can estimate after how much time $\tilde{t}$ in the average rest frame two electrons, initially at rest at a distance $\tilde{d}$ from each other, are at a distance, e.g., of $1.1\tilde{d}$, when being accelerated by a constant repulsive force $|\langle \tilde{\boldsymbol{F}}_C \rangle|$. Simple non-relativistic kinematics implies that $\tilde{t} \sim \sqrt{m\tilde{d}/10/|\langle \tilde{\boldsymbol{F}}_C \rangle|}$. In the laboratory frame, $\tilde{t}$ is di-

lated by a factor $\gamma$ and during this time the two electrons propagate along the longitudinal direction over a distance $l_\parallel \sim \gamma\tilde{t}$. By estimating $\tilde{d} \sim \tilde{r}'$, we obtain that for the parameters used for Fig. 2, $l_\parallel \sim 3$ mm and the experimental results, e.g., in [5] show that distances between the electron source and a strong laser field much smaller than 300 $\mu$m are achieved. However, in the case of the second numerical example reported in Fig. 3, a similar estimation provides $l_\parallel$ equal to 80 $\mu$m for the panel a) (which is still acceptable), equal to 8 $\mu$m for the panel b), equal to 3 $\mu$m for the panel c), and, finally, equal to 2 $\mu$m for the panel d). Notice that while with the parameters of Fig. 2 and Fig. 3 the longitudinal component of the force $\tilde{F}_{C,\parallel}$ is typically many orders of magnitude smaller than each of the transverse components, in the panels c) and d) it is actually of the same order of magnitude. This still implies that the parallel displacement between the electrons $\tilde{d}_\parallel \sim \tilde{r}'_\parallel$ increases by 10% after a propagation in the laboratory frame of 3 $\mu$m for panel c), but for panel d) this happens already after 0.6 $\mu$m.

Finally, we observe that it is possible, via the Ehrenfest's theorem, to relate this classical estimate to the evolution of the distance between the centers of the wave packets.

---



\end{document}